
\input harvmac


\Title{\vbox{\baselineskip12pt
\hbox{McGill/95-38}
\hbox{CERN-TH/95-213}
\hbox{hep-th/9508045}
}}
{\vbox{\centerline{Dynamics of Extreme Black Holes}
\smallskip     \centerline{and}
\smallskip     \centerline{Massive String States}}}

\centerline{Ramzi R.~Khuri$^{1,2}$ and Robert C.~Myers$^1$}
\bigskip\centerline{$^1${\it Physics Department, McGill
University, Montreal, PQ, H3A 2T8 Canada}}
\bigskip\centerline{$^2${\it CERN,
 CH-1211, Geneva 23, Switzerland}}
\vskip .3in
In a recent paper, Duff and Rahmfeld argued that certain
massive $N_R=1/2$ states of the
four-dimensional heterotic string correspond to extreme
black hole solutions. We provide further,
dynamical, evidence for this identification by comparing the
scattering of these elementary string states with that
of the corresponding extreme black holes, in the limit
of low velocities.

\vskip .3in
\Date{\vbox{\baselineskip12pt
\hbox{August 1995}}}

\def\sqr#1#2{{\vbox{\hrule height.#2pt\hbox{\vrule width
.#2pt height#1pt \kern#1pt\vrule width.#2pt}\hrule height.#2pt}}}

\def\a{\alpha}
\def\z{\zeta}
\def\b{\beta}
\def\p{\a}
\def\q{Q}
\def\ie{{\it i.e.,}\ }
\def\eg{{\it e.g.,}\ }
\def\st{\sqrt{3}}
\def\E{\tilde{E}}
\def\xch{\zeta}
\def\zz#1#2{\z^{#1R}\!\cdot\!\z^{#2R}}
\def\cc#1#2{\xch^{#1L}\!\cdot\!\xch^{#2L}}

\lref\dufr{M.~J.~Duff and J.~Rahmfeld, Phys. Lett.
{\bf B345} (1995) 441.}

\lref\dufkmr{M.~J.~Duff, R.~R.~Khuri, R.~Minasian and
J. Rahmfeld, Nucl. Phys. {\bf B418} (1994) 195.}

\lref\gros{D.~J.~Gross and J.~Sloan, Nucl. Phys. {\bf B291}
(1987)  41.}

\lref\monscat{R.~R.~Khuri, Phys. Lett. {\bf B294} (1992) 331.}

\lref\stscat{R.~R.~Khuri, Phys. Lett. {\bf B307} (1993) 302.}

\lref\fbscat{A.~G.~Felce and T.~M.~Samols, Phys. Lett.
{\bf B308} (1993) 302.}

\lref\macstr{R.~R.~Khuri, Nucl. Phys. {\bf B403} (1993) 335.}

\lref\hmono{R.~R.~Khuri, Phys. Lett. {\bf B259} (1991) 261;
Nucl. Phys. {\bf B387} (1992) 315.}

\lref\multhmono{R.~R.~Khuri, {\it Solitons and Instantons in
String Theory}, Princeton University Doctoral Thesis, October
1991, UMI-92-025-32, {\bf 52 08B}, February 1992.}

\lref\manton{N.~S.~Manton, Phys. Lett. {\bf B110} (1982) 54;
Phys. Lett. {\bf B154} (1985) 397;
Phys. Lett. {\bf B157} (1985) 475.}

\lref\atiyah{M.~F.~Atiyah and N.~J.~Hitchin, Phys. Lett.
{\bf A107} (1985) 21; {\it The Geometry and Dynamics of
Magnetic Monopoles}, (Princeton University Press, 1988).}

\lref\fere{R.~C.~Ferrell and D.~M.~Eardley,
Phys. Rev. Lett. {\bf 59} (1987) 1617;
in {\it Frontiers in numerical relativity,}
Eds. C.R.~Evans, L.S.~Finn and D.W.~Hobill
(Cambridge University Press, 1988).}

\lref\schsen{J.~H.~Schwarz and A.~Sen, Phys. Lett. {\bf B312}
(1993) 105; A.~Sen, Int. J. Mod. Phys. {\bf A9} (1994) 3707.}

\lref\ghs{D.~Garfinkle, G.~T.~Horowitz and A.~Strominger,
Phys. Rev. {\bf D43} (1991) 3140.}

\lref\stata{A.~Papapetrou, Proc. R. Irish Acad. {\bf A51}
(1947) 191; S.D. Majumdar, Phys. Rev. {\bf D35} (1947) 930.}

\lref\senb{A.~Sen, Nucl. Phys. {\bf B440} (1995) 421.}

\lref\hw{C.~F.~E.~Holzhey and F.~Wilczek, Nucl. Phys.
{\bf B360} (1992) 447.}

\lref\salam{S.~W.~Hawking, Monthly Notices Roy. Astron. Soc.
 {\bf 152} (1971) 75; A.~Salam in {\it Quantum Gravity: an
Oxford Symposium} (Eds. Isham, Penrose and Sciama, O.U.P. 1975);
G.~'t Hooft, Nucl. Phys. {\bf B335} (1990) 138.}

\lref\shirscat{K.~Shiraishi, Nucl. Phys. {\bf B402} (1993) 399;
``Many-body systems in Einstein-Maxwell-Dilaton theory,''
preprint
gr-qc/9507029.}

\lref\morebranes{R.~R.~Khuri and R.~C.~Myers, in preparation.}

\lref\shir{K.~Shiraishi, J. Math. Phys. {\bf 34} (1993) 1480.}

\lref\kawai{H.~Kawai, D.~C.~Lewellen and S.~-H.~H.~Tye,
Nucl. Phys. {\bf B269} (1986) 1.}

\lref\ghmrii{D.~J.~Gross, J.~A.~Harvey, E.~Martinec and
R.~Rohm, Nucl. Phys. {\bf B267} (1986) 75.}

\lref\jerome{J.~P.~Gauntlett, J.~A.~Harvey, M.~M.~Robinson
and D.~Waldram, Nucl. Phys. {\bf B411} (1994) 461.}

\lref\schwarz{J.~H.~Schwarz, Phys. Rep. {\bf 89} (1982) 223.}

\lref\gibbons{G.~W.~Gibbons, Nucl. Phys. {\bf B207} (1982) 337.}

\lref\ght{G.~W.~Gibbons, G.~T.~Horowitz and P.~K.~Townsend,
Class. Quant. Grav. {\bf 12} (1995) 297.}

\lref\senbh{A.~Sen, ``Extremal black holes and elementary string
states'', preprint hep-th/9504147.}

\lref\peet{A.~Peet, ``Entropy and supersymmetry of $D$
dimensional extremal electric black holes versus string states'',
preprint hep-th/9506200.}

\lref\gsw{M.~B.~Green, J.~Schwarz and E.~Witten,
{\it Superstring Theory}, (Cambridge University Press, 1987),
Vol. 1.}


\newsec{Introduction}

A particle whose mass is greater than the Planck mass
possesses a Schwarzschild radius that is greater than its
Compton wavelength. As a result even within quantum
mechanics, such a particle may exhibit
an event horizon and thus behave like a black hole \salam.
It seems, however, that testing this conjecture requires
an understanding of quantum gravity.
Since string theory is a candidate theory of quantum gravity
and it predicts the existence of states with Planck scale masses,
it should provide a consistent framework in which the above
conjecture
might be tested in a concrete manner. Recently
the suggestion was made that certain massive excitations of
four-dimensional superstrings are indeed black holes \dufkmr.
In subsequent work \dufr, this claim was supported by
the discovery that certain extreme black holes
could be identified with electrically charged states
in the Schwarz-Sen spectrum, which in turn can be identified
with elementary string states \schsen.
On the basis of $S$-duality, the corresponding solitonic
magnetically
charged spectrum conjectured by Schwarz and Sen \schsen\
would also be described by extreme black holes \dufr.

In this paper, we provide dynamical evidence in support
of this conjecture by comparing the tree-level scattering
amplitudes of the massive string states of the Schwarz-Sen
spectrum discussed in \dufr\ with the computation of the
low velocity scattering of the corresponding extreme black holes
discussed in \dufkmr.
In section 2, we summarize the main features of the above
conjecture as presented in \dufr. In section 3, we summarize
the results of Shiraishi \shir\ using Manton's prescription
\manton\ for the computation of the metric on moduli space
for the extreme black holes. In section 4, we compute the
scattering amplitudes of the corresponding massive string
states. We conclude with a discussion in section 5.

\newsec{Massive String States as Extreme Black Holes}

Consider the four-dimensional heterotic string theory arising
from toroidal compactification. For a generic point in the
moduli space, the low energy effective theory is $N=4$
supergravity
coupled to twenty-two abelian vector multiplets. Thus the
bosonic fields include the metric, the dilaton, the axion,
$28$ abelian gauge fields, and $132$ scalar moduli fields.
This theory then contains a rich array of black hole solutions
with electric and/or magnetic charges \senb.
Of the static charged solutions, a special subset have been
identified
as involving a single $U(1)$ vector field and a single scalar,
as well as
the metric \dufkmr. These solutions can then be regarded as
solutions for a theory described by
\eqn\mbone{
I= {1\over 16\pi G}\int d^4\!x\,\sqrt{-g}\left [R-
{1\over 2}(\partial
\phi)^2-{1\over 4}e^{-a \phi}F^2 \right]\ .}
Here, $F$ is a linear combination of
the field strengths and their duals appearing in the
original effective action, while $\phi$ represents an
appropriate linear combination of the dilaton and the moduli
fields. Within field theory, static black hole
solutions have been found
for arbitrary $a$ by Gibbons \gibbons. In the context of
low energy heterotic string theory,
solutions have been identified which correspond to
fixing the above
scalar-Maxwell coupling to $a=0,$ $1/\sqrt{3},$ $1,$ or
$\sqrt{3}$.
(We will assume $a\ge0$ without loss of generality in the
following.)
The case $a=0$ yields the Reissner-Nordstrom solution, for
which one actually has $\phi=constant$, and was recognized to be
a solution of string theory in \dufkmr.
The dilaton black hole \ghs\ arises for $a=1$.
The $a=\st$ case corresponds to the
Kaluza-Klein black hole \dufkmr\ and the ``H-monopole''
black hole \refs{\dufkmr,\hmono},
which are related to each other by $T$-duality.
The $a=1/\sqrt{3}$ black hole
has also been more recently recognized to be a solution of string
theory \ght.

In the following, we will be interested in electrically
charged extreme black holes, since only extreme black
holes possess zero Hawking temperature and thus may be candidates
for identification with elementary particles.
 For spherically symmetric
black holes (\ie with vanishing angular momentum), the extreme
solutions for the action \mbone\ satisfy
\eqn\xtremity{(Gm)^2={Q^2\over 4(1+a^2),}}
where $m$ is the ADM mass, and $Q$, the electric charge where
$Q\equiv\oint_\infty e^{-a\phi}{}{\tilde F}/4\pi$ (where the
``tilde'' denotes the Hodge dual).
For such extreme black holes with like charges, the repulsive
electrostatic force is precisely balanced by the attractive
static forces of the gravitational and scalar fields.
Hence it is possible to construct
static multi-extreme black hole solutions for the action \mbone.
These have long been known for the Reissner-Nordstrom case (\ie
$a=0$) \stata, and also for the dilaton black holes (\ie
$a=1$) \ghs\ and the ``H-monopoles'' \multhmono.
Shiraishi \shir\ found multi-extreme black hole
solutions for arbitrary $a$, but recall only the values of
$a=0, (1/\st), 1, \st$
have so far been demonstrated to arise from heterotic string
compactifications.
The precise cancellation of forces no longer occurs when the
extreme black holes are in motion, and so they will
scatter in a nontrivial way. The latter will be the focus of
the discussion in section 3.

Within the full four-dimensional string theory,
Schwarz and Sen produced a spectrum of states satisfying a
certain
Bogomol'nyi bound, which is invariant under both $O(6,22; Z)$
and
$SL(2,Z)$ \schsen. Of these, a subset can correspond to
elementary string states. The latter carry only electric
charge in which case the Bogomol'nyi bound can be reduced to
\refs{\schsen,\dufr}
\eqn\mbseven{(Gm)^2={1\over 16}\q^a (I+L)_{ab} \q^b=
 {1\over 8}(\q_R)^2}
where the right- and left-projections of the charge vector
are defined as $\q_R = {1\over 2}(I+L) \q$ and
$\q_L = {1\over 2} (I-L) \q$ with $L$ the invariant
metric on $O(6,22)$, and ${\q}^a$ the charges of the 28 abelian
gauge fields. For elementary string states, $\q_{L,R}$ are
related to the internal momenta of the left- and right-movers.

The mass of a heterotic string state in the Neveu-Schwarz sector
(which is degenerate with the Ramond sector) is given
by\footnote{$^1$}{We adopt the normalization ${\a}^\prime
=2$ throughout the paper.}
\eqn\mbeight{m^2= \left(\alpha_R \right)^2 +2 N_R -1 =
 \left(\alpha_L \right)^2 +2 N_L -2\ ,}
where $N_{L,R}$ are the left- and right-oscillator numbers,
and $\a_{L,R}$ are the internal momenta for the left- and
right-moving
modes. Given $\q_{L,R}^2=8G^2\,\a_{L,R}^2$,
a comparison of \mbseven\ and \mbeight\ shows that the string
states
satisfying the Bogomol'nyi bound have $N_R=1/2$. Given the
correspondence
of the charge vector with the internal momenta, as well as the
masses,
these elementary massive $N_R=1/2$ string states and their
superpartners have the correct quantum number
to fit into the Schwarz-Sen spectrum \schsen.

In more recent work, Duff and Rahmfeld \dufr\ show that a
subset
of these $N_R=1/2$ states in the Schwarz-Sen spectrum may
also be identified as the extreme limits of certain black hole
solutions.
For these solutions, the low-energy string action can be
truncated to \mbone\ and the scalar-Maxwell parameter is given
by
$a=\sqrt 3$ for $N_L=1$ and $a=1$ for $N_L>1$ and $\a_L=0$.
In particular, the charge vector $\q^a=2\sqrt{2}G\,\delta^{a,1}$
was
shown to correspond to an $a=\sqrt{3}$ black hole.
This choice satisfies $\q_L^2=\q_R^2$
and hence $N_L=1$ in \mbeight. From \mbseven\ it follows that
the mass is given
by $m^2=1/2=Q^2/16G^2$, which coincides with
\xtremity\ in the extreme limit \dufr. $O(6,22)$ transformations
and rescaling the charge vector then yield all other charge
vectors satisfying $Q_L^2=Q_R^2$, and hence all
$N_R=1/2$, $N_L=1$ states should
correspond to $a=\st$ black holes.
Similarly, a particular $a=1$ black hole was shown to correspond
to $\q^a=2\sqrt{2}G(\delta^{a,1}+\delta^{a,7})$. In this case,
one
has $m^2=2=Q^2/8G^2$ which coincides with \xtremity\ in the
extreme limit for $Q^2=16G^2$. In this case,
$\q_L=0$. Again, $O(6,22)$ transformations
and rescaling the charge vector will yield all charge
vectors satisfying these two conditions, and hence all
$N_R=1/2$, $N_L>1$
and $\a_L=0$ states correspond to $a=1$ black holes.
Other states with $N_L>1$ and $\a_L\ne0$
should also be extreme black holes, but a
truncation to an effective action of the form \mbone\ is not
possible. Note that neither $a=0$ \dufr\
nor $a=1/\sqrt{3}$ extreme black holes belong to the
spectrum. Furthermore, the $a=1$ extreme dilaton black holes of
Ref.~\ghs\
also do not belong to the spectrum \dufr.

\newsec{Extreme Black Holes in Slow Motion}

The existence of static multi-soliton solutions, of which
the multi-extreme black hole solutions
described in the previous section are examples,
 relies on the cancellation
of the scalar, vector and tensor exchange forces
(the so-called ``zero-force''
condition). If the solitons are given velocities, however,
the zero-force condition ceases to hold and dynamical,
velocity-dependent forces arise. The full time-dependent
equations
of motion that result are highly nonlinear and in general very
difficult to solve. In the absence of exact time-dependent
multi-soliton solutions, Manton's method \manton\
for the computation of the metric on moduli space represents
a good low-velocity approximation to the exact dynamics of the
solitons. Manton's prescription for the study of soliton
scattering
may be summarized as follows: One begins with a static
multi-soliton solution, and gives the moduli characterizing
this configuration a time-dependence. One then finds $O(v)$
corrections to the fields by solving the constraint
equations of the system with time-dependent moduli.
The resultant time-dependent field configuration only
satisfies the full time-dependent field equations to lowest
order in the velocities, but provides
an initial data point for the fields and their time derivatives.
Another way of saying this is that the initial motion is
tangent to the set of exact static solutions. An effective
action
describing the motion of the solitons is determined by
replacing the solution to the constraints into the field theory
action. The kinetic action so obtained
defines a metric on the moduli space of static solutions, and
the geodesic
motion on this metric determines the dynamics of the solitons.
This approach was first applied to study the scattering of
BPS monopoles \manton, and a complete calculation of the
corresponding
moduli space metric and a description of its geodesics was
worked out by Atiyah and Hitchin \atiyah.

Manton's method was subsequently adapted to general
relativistic actions by Ferrell and Eardley \fere\ for the
low-energy scattering of extreme Reissner-Nordstrom black holes.
More recently, Shiraishi \shirscat\ adapted the method of
Ref.~\fere\ to obtain the metric on moduli space for generalized
multi-black hole solutions of the action \mbone. The resulting
effective Lagrangian describing the interactions of $N$
extremally charged black holes is
\eqn\effact{
\eqalign{L=&-\sum_{i=1}^N m_i + \sum_{i=1}^N {1\over 2}
m_i v_i^2\cr
&+{3-a^2\over16\pi}\int d^3x F(x)^{2(1-a^2)/(1+a^2)}
\sum_{i\ne j}^NG
m_im_j|\vec{v}_i-\vec{v}_j|^2{\vec{r}_i\cdot\vec{r}_j
\over r_i^3\, r_j^3}
\cr}}
where $\vec{r}_i=\vec{x}-\vec{x}_i$, and $\vec v_i$ and
$\vec x_i$ are,
respectively, the velocity and position of the $i$'th black hole.
Also,
\eqn\efff{
F(x)=1+(1+a^2)\sum_{i=1}^N{Gm_i\over r_i}\ .
}
As usual in the Manton method, this effective action represents
the leading terms up to $O(v^2)$ in a small velocity expansion,
and
so neglects the effects of any radiation fields which would only
contribute
at higher orders. The first two terms in \effact\ correspond to
the expected free particle Lagrangian to $O(v^2)$. The remaining
term is the interaction Lagrangian which as expected vanishes as
the relative velocities go to zero. In general, this contribution
is highly nonlinear involving up to $N$-body interactions.
Collecting all of the $O(v^2)$ terms yields a metric on the
moduli space of these $N$ black hole configurations.

The above interactions simplify for two values of the
scalar-Maxwell coupling,
$a=\st$ and $a=1$, which are precisely the values of interest in
the present paper. The effective Lagrangian for extreme $a=\st$
black holes
reduces to the free terms only \refs{\shir,\shirscat}
\eqn\lagtriv{L=-\sum_{i=1}^N m_i + \sum_{i=1}^N {1\over 2}
m_i v_i^2\ .}
In other words,
the leading order velocity-dependent (\ie $O(v^2)$)
dynamical force between the black holes is zero, and the
low-velocity scattering is trivial. Thus one infers
that the metric on the moduli space of these
$a=\st$ extreme black holes is flat. A similar flat metric has
been found for $H$-monopoles \monscat,  fundamental strings
\stscat\ and $D=10$ fivebranes \fbscat. In fact, one can show
that
a flat metric describes the motion of all $\kappa$-symmetric
$p$-branes \morebranes.

For $a=1$, the Lagrangian \effact\ simplifies in that it only
involves two-body interactions. Thus the effective
Lagrangian is easily determined to be simply \shirscat
\eqn\kinlag{L=-\sum_{i=1}^N m_i + \sum_{i=1}^N {1\over 2}
m_i v_i^2 + {1\over 2} \sum_{i\ne j}^N Gm_im_j
{|\vec v_i - \vec v_j|^2
\over |\vec x_i-\vec x_j|\hphantom{^2}}\ .}
In this case, the nontrivial interaction
leads to a center-of-mass deflection angle
for the scattering of two $a=1$ black holes:
$\theta=2\tan^{-1}(GM/b)$, where $M=m_1+m_2$ is the total mass
and $b$ is the impact parameter \shirscat. The resulting
differential
cross section then has the Rutherford scattering form \shirscat
\eqn\rutherford{{d\sigma\over d\Omega}={1\over 4}
{(GM)^2\over \sin^4(\theta/2)}.}

Note that while Shiraishi's work describes the dynamics of black
hole solutions for the truncated action \mbone, we are
interested in
black hole solutions of the low energy effective string action.
In this case,
we must consider the possibility that the time-dependent
solutions
will involve contributions from the other bosonic fields in the
full theory. It is not hard to show that the motion of the black
holes does not induce nonvanishing contributions from the other
moduli scalars or $U(1)$ vectors. Similarly for $a=\st$, the
axion remains
vanishing. On the other hand for $a=1$, the motion of the black
holes
will lead to a nontrivial axion. The effective action which must
be considered then is
\eqn\mbtwo{
\eqalign{
I= {1\over 16\pi G}\int d^4\!x\,\sqrt{-g}&\left [R-{1\over 2}
(\partial
\phi)^2-{1\over 4}e^{- \phi}F^2\right.\cr
&\ \left.-{1\over 2}e^{2\phi}(\partial\rho)^2+{1\over 8}
\rho\,\varepsilon^{abcd}F_{ab}F_{cd}\right]\cr}}
where $\rho$ is the scalar axion field. The equation of motion
for the
latter is
\eqn\axion{
\nabla^a(e^{2\phi}\nabla_a\rho)=-{1\over8}
\varepsilon^{abcd}F_{ab}F_{cd}
=\vec E\cdot\vec B}
where $\vec E$ and $\vec B$ are the electric and magnetic
fields, respectively,
as the spatial three vectors. Now for static electrically
charged
black holes, the absence of any magnetic fields means that the
axion
is consistently set to $\rho=0$.
When these charged black holes are in relative motion
though, the presence of both electric and magnetic fields
requires that
a nonvanishing axion field for a consistent solution.

However, we will now argue that effective Lagrangian \kinlag\
 remains
valid as a leading approximation
to the full results for stringy black holes interacting at
large separations.
For small relative velocities, the magnetic field will be $O(v)$
and so by \axion, the axion field is of the same order.
The key observation is that
since the magnetic fields are proportional to the electric
charges,
and hence the black hole masses by \xtremity, the induced axion
field is proportional to products of masses, $G^2m_im_j$.
Thus when the axion field is substituted
back into the action \mbtwo, following the Manton method, the
resulting
interactions schematically take the form $G^3m^4 v^2/r^3$
(at least to leading order). Here, the $1/r^3$
dependence on the black hole separations is inferred by
dimensional analysis. Similarly the modifications of the metric,
dilaton and gauge fields induced by the axion can only lead to
modifications
of the same or higher order in $Gm/r$. Thus if we consider
making an expansion
of the interaction lagrangian in $Gm/r$ for black holes
interaction
at large separations, we see that Shiraishi's
results \kinlag\ remain the leading contribution, and that the
axion will
only modify the higher order interactions.
Our calculations of string scattering
will only be sensitive to the leading $O(1/r)$ interaction, and
hence the above results are sufficient for our purposes.

\newsec{Amplitudes for Massive String States}

We now consider the scattering of
massive heterotic string states characterized by:
(i) $N_R={1\over2}$,
$N_L=1$ and hence $\a_L^2=\a_R^2$; and (ii) $N_R={1\over2}$,
$N_L>1$ and $\a_L=0$. These, of course, are expected to
correspond to the $a=\sqrt{3}$ and $a=1$ extreme black holes
as discussed above.
We will calculate the corresponding four-point closed string
amplitudes
using the
methods of Ref.~\kawai, where it was shown that the left- and
right-moving
world-sheet modes could be treated completely independently.
The final closed string
amplitudes are constructed by sewing together two open string
amplitudes for these independent modes. The four particle
interactions
described by the full amplitudes include
contributions from all possible massive string states, as well
as
the exchange of the massless fields appearing in the low-energy
effective action (\ie gravitons, vectors and scalars).
The results of section 3 for the scattering of extreme black
holes only
account for the latter massless particle exchange. The former
results
also only describe the scattering of black holes corresponding
to a fixed embedding of the fields in \mbone\ into
the full low energy string theory. Of course, the previous
analysis
was also limited to low velocity scattering. Hence in the
string amplitude, we choose particles 1 and 4 to be identical
string
states with the same internal momentum and polarizations, and
we make a
similar choice for particles 2 and 3. Then we arrange the
spacetime kinematics of these particles to yield low velocity
elastic scattering. Finally the string amplitude will be
examined for contributions with a $t$-channel pole, which will
correspond to the exchange of massless particles.

Following the procedure of \kawai\ we split our vertex
operators into left- and
right-movers. Since our string states are all in the
Neveu-Schwarz sector of the right movers with $N_R=1/2$, the
right-moving excitations can be considered as a massless
superstrings, in a ten-dimensional sense. Similarly
the left-states correspond to (twenty-six-dimensional)
bosonic strings with masses $M_L^2=2(N_L-1)$.
For each state then, there is a left- and right-momentum vector
with 26 and 10 components, respectively. In accord with
our choice of kinematics,
$p^\mu_{1L,R}$ and  $p^\mu_{2L,R}$ are the incoming momenta and
$-p^\mu_{3L,R}$ and $-p^\mu_{4L,R}$ are the outgoing momenta.
Hence $p^\mu_1+p^\mu_2+p^\mu_3+p^\mu_4=0$ for both left- and
right-momenta.
The spacetime four-momenta, which we denote $k^\mu_i$,
correspond to the
first four components of either $p^\mu_{iL,R}$.
Hence we have:
\eqn\momenta{\eqalign{
p^\mu_{1L,R}&=(\ \  E, \ \ E\vec v_1,\ \ \p_{1L,R}),\cr
p^\mu_{2L,R}&=(\ \ \E,   -\E\vec v_2,\ \ \p_{2L,R}),\cr
p^\mu_{3L,R}&=(   -\E,\ \ \E\vec v_3,   -\p_{2L,R}),\cr
p^\mu_{4L,R}&=(    -E,    -E\vec v_4,   -\p_{1L,R}),\cr}}
where $\vec{v}_i$ are the three-dimensional spatial
velocities, and $\p_{iL,R}$ are the internal
left- and right-momenta, respectively. We have used
energy-momentum
conservation to write $E_1=E_4=E$ and $E_2=E_3=\E$, and further
we have
$v_1^2=v_4^2=v^2$ and $v_2^2=v_3^2=w^2$. For comparison to
the low velocity scattering in the previous section, we
restrict
$v^2,w^2<<1$.
Working in the center-of-mass frame in the four-dimensional
spacetime, $E\vec v_1=\E\vec v_2$ and $E\vec v_4=\E\vec v_3$.
Note that $\vec{v}_1\cdot\vec{v}_4=v^2 \cos\theta$ and
$\vec{v}_2\cdot\vec{v}_3=w^2 \cos\theta$,
where $\theta$ is the scattering angle in the center-of-mass
frame.
Further, one has
\eqn\enmom{E_i^2(1-v_i^2)=m_i^2=\p_{iR}^2=\p_{iL}^2 +
2(N_{iL}-1)\ .}
Note that for slow motion scattering with $v_i^2<<1$, one
has $E_i=m_i+{1\over2}m_iv_i^2+O(v_i^4)$.

We choose all
the internal momenta to be parallel (\ie $\p_{2R}=\beta\p_{1R}$
and $\p_{2L}
=\beta\p_{1L}$). This ensures that our
string states all couple to the scalars and gauge
fields in the same way. Note that this imposes $m_2=\beta m_1$
and $(N_{2L}-1)=\b^2(N_{1L}-1)$.
Similarly to compare to elastic scattering of
identical black holes, we must choose the polarization tensors
to be the same for states 1 and 4, as well as for 2 and 3
(\ie $\zeta^1=\zeta^4$ and $\zeta^2=\zeta^3$.)
These polarization tensors are schematically
represented as tensor products of separate
polarization tensors for the left- and right-movers
(\ie $\zeta^i=\zeta^{iR}\otimes\xch^{iL}$), which satisfy
$p_{iR}\cdot\zeta_{iR}=0=p_{iL}\cdot\xch_{iL}$. For comparison
with
spherically symmetric black holes, the string states should be
scalars
in the four-dimensional spacetime. The simplest choice which
satisfies this
restriction, and which we make below, is to choose the
polarization
tensors only to take nonvanishing values for the internal
directions.

It is useful to write out the higher-dimensional
Mandelstam variables for the above configuration.
For the right-movers,
\eqn\mandelstamr{\eqalign{
s_R&=-(p_{1R} + p_{2R})^2=(E+\E)^2-(1+\b)^2\p_{1R}^2=O(v^2),\cr
t_R&=-(p_{2R} + p_{3R})^2=-2E^2v^2(1-\cos\theta),\cr
u_R&=-(p_{1R} + p_{3R})^2=
(E-\E)^2-2E^2v^2(1+\cos\theta)-(1-\b)^2\p_{1R}^2=O(v^2)
\cr}}
with $s_R+t_R+u_R=0$.
Similarly, for the left-movers,
\eqn\mandelstaml{\eqalign{
s_L=-(p^\mu_{1L} + p^\mu_{2L})^2=&(E+\E)^2-(1+\b)^2\p_{1L}^2\cr
=&2(1+\b)^2(N_{1L}-1)+O(v^2),\cr
t_L=-(p^\mu_{2L} + p^\mu_{3L})^2=&-2E^2v^2(1-\cos\theta),\cr
u_L=-(p^\mu_{1L} + p^\mu_{3L})^2=&
(E-\E)^2-2E^2v^2(1+\cos\theta)-(1-\b)^2\p_{1L}^2\cr
=&2(1-\b)^2(N_{1L}-1)+O(v^2),
\cr}}
with $s_L+t_L+u_L=4(1+\b^2)(N_{1L}-1)$.
Note that the spacetime Mandelstam variable $t$ coincides with
that
of either the left- or right-momenta
(\ie $t=-(k_2+k_3)^2=t_R=t_L$).
On the other hand in general, the spacetime Mandelstam variables
$s$ and $u$ need not coincide with those of the left- and
right-momentum vectors. For later purposes, we note that
\eqn\mandelstam{\eqalign{
s_R&=-(k_1+k_2)^2-(1+\b)^2\a_R^2=s-(1+\b)^2m_1^2\cr
s_L&=-(k_1+k_2)^2-(1+\b)^2\a_L^2=s_R+2(1+\b)^2(N_{1L}-1)\cr
u_R&=-(k_1+k_3)^2-(1-\b)^2\a_R^2=u-(1-\b)^2m_1^2\cr
u_L&=-(k_1+k_3)^2-(1-\b)^2\a_L^2=u_R+2(1-\b)^2(N_{1L}-1).\cr}}

Following the methods of Ref.~\kawai\ (see also \gsw), the
four-point amplitude for heterotic string states can be
expressed in terms of four-point amplitudes for the
open left-moving bosonic and open right-moving supersymmetric
states:
\eqn\ampone{A_{4,het}\approx
\sin(\pi t_R/2)\,A_{4,sup}(t_R, u_R)\, A_{4,bos}(s_L, t_L),}
where $A_{4,sup}(t_R, u_R)$ corresponds to a
$t$-$u$ channel open superstring amplitude, and
$A_{4,bos}(s_L, t_L)$ corresponds to an $s$-$t$ channel open
bosonic string amplitude. Note that the term $\sin(\pi t_R/2)$,
which is needed to ``sew'' the two open string amplitudes, could
equally well have been written as $\sin(\pi t_L/2)$,
since $t_L=t_R$ for the present kinematics.
In fact this would be possible in general since it is always
the case
that the difference between corresponding right- and
left-momentum Mandelstam variables is always an integer
multiple of $8$. Hence the sewing factor is always
identical no matter from which side the momenta are chosen.
In the small velocity limit, the sewing term is of
$O(v^2)$ for all the cases we consider, and we may write
$\sin(\pi t_R/2)\approx \pi t_R/2$.

For the right-moving superstring contribution, one has
\eqn\susyamp{
A_{4,sup}(t_R, u_R)={\Gamma(-t_R/2)\Gamma(-u_R/2)
\over \Gamma(1+s_R/2)} K_R(p_{iR}, \zeta^{jR})}
where $K_R$ is the kinematic factor for four massless
superstring vector states. Recall that for simplicity,
we will choose the
polarization vectors to point in the internal directions.
Further note that since the internal momentum vectors
are all parallel, one has $p_{iR}\cdot\zeta^{jR}=0$ for any
$i$ and $j$. This greatly simplifies the analysis since
in this case the kinematic factor reduces to
\eqn\simpamp{
K_R=-{1\over4}(s_R\,t_R\,\zz 1 3\,\zz 2 4
+s_R\,u_R\,\zz 2 3\,\zz 1 4+t_R\,u_R\,\zz 1 2\,\zz 3 4)\ .}
Note that this factor is $O(v^4)$ since $s_R$, $t_R$ and $u_R$
are each $O(v^2)$ --- see eq.~\mandelstamr. The Gamma
functions in \susyamp\ give poles in $t_R$ and $u_R$, and
so the net result is that $A_{4,sup}(t_R, u_R)$ is of order $1$.

Combining the superstring and sewing factors, one has
\eqn\susy{
\eqalign{
\sin(\pi t_R/2)\,A_{4,sup}(t_R, u_R)
\approx-{\pi\over2}&\left(
{s_R\,t_R\over u_R}\,\zz 1 3\,\zz 2 4
+s_R\,\zz 2 3\,\zz 1 4 \right.\cr
&\left.\vphantom{s_R\,t_R\over u_R}+t_R\,\zz 1 2\,\zz 3 4\right)
\ .\cr}}
This final result is $O(v^2)$, thus if the total amplitude is
to be $O(1)$, $A_{4,bos}$ must supply an $O(v^{-2})$ pole.
Further if we are to identify a $t$-channel
pole, it must arise as a $1/t_L=1/t_R$ factor
in $A_{4,bos}$. The analysis of the bosonic string
factor differs for the two cases under consideration and
so we separate the discussion:

\bigbreak

\noindent{{\bf (i)} $N_R={1\over2}$ and $N_L=1$}

With $N_L=1$ the left-movers also correspond to four massless
vectors (in a twenty-six-dimensional sense).
In this case, a number of simplifications occur for the kinematic
variables. In particular since $N_{1L}=N_{2L}=1$,
we see from eq.~\mandelstaml\ that $s_L=O(v^2)=u_L$ (and
$s_L+t_L+u_L=0$). The amplitude then reduces to
\eqn\bosamp{
A_{4,bos}(s_L, t_L)={\Gamma(-s_L/2)\Gamma(-t_L/2)
\over \Gamma(1+u_L/2)}\, \tilde{K}_L(p_{iR}, \zeta_{jR})}
where $\tilde{K}_L$ is the appropriate bosonic string kinematic
factor. Again all of the left
internal momentum vectors are parallel, and so with internally
pointing polarization vectors,
$(p_{iR}\cdot\xch^{jL})=0$ for any $i$ and $j$.
With these conditions the kinematic factor takes the same form
as for the superstring with
\eqn\simpampa{
\tilde{K}_L=-{1\over4}(s_L\,t_L\,\cc 1 3\,\cc 2 4
+s_L\,u_L\,\cc 2 3\,\cc 1 4+t_L\,u_L\,\cc 1 2\,\cc 3 4)\ .}
Note that this factor is again $O(v^4)$ while the Gamma
functions in \bosamp\ contribute poles in $s_L$ and $t_L$, each
of $O(v^{-2})$. Thus just as for the right-movers, the
net result is $O(1)$.

This leaves the total heterotic string amplitude as $O(v^2)$,
which indicates that there is no scattering
of these string states, to leading order
in the low velocity approximation.  Since a flat metric
was found for the moduli space of $a=\st$ extreme black holes,
the result supports the identification of these string states
and $a=\st$ black holes.

\bigbreak

\noindent{{\bf (ii)} $N_R={1\over2}$, $N_L>1$ and $\a_L=0$}

With $N_L>1$, the left-movers also correspond to four massive
tensor states (in the twenty-six-dimensional sense). In the
following,
we will assume that $N_{2L}\ge N_{1L}$ and hence $\beta\ge1$.
Explicit results for the
required amplitudes are not readily available in the literature,
however it is not difficult to perform the necessary
calculations.
We represent the left-moving states with the vertex operators
\eqn\leftone{
V_i=\xch^{iL}_{\a\cdots\b}\,\partial X^\a\! \cdots
\partial X^\b e^{ip_{iL}\cdot X}}
where $\xch^{1L}=\xch^{4L}$ and $\xch^{2L}=\xch^{3L}$ are
symmetric, traceless tensors
with $N_{1L}$ and $N_{2L}$ indices, respectively.
Again both polarization tensors only take
nonvanishing values for internal directions and since
$\a_{iL}=0$,
the contraction of any of the momenta $p_{iL}$ with any indices
on the polarization tensors vanishes. This simplifies
the corresponding amplitude, and schematically one finds
\eqn\factora{\eqalign{
A_{4,bos}\approx&
\int\prod_{i=1}^4 dx_i {|x_a-x_b||x_a-x_c||x_b-x_c|\over
dx_a\,dx_b\,dx_c}
\prod_{i>j}|x_i-x_j|^{p_{iL}\cdot p_{jL}}
\left[{1\over(x_2-x_3)^2}\right]^{N_{2L}-N_{1L}}\cr
&\left[{1\over(x_1-x_2)^2(x_3-x_4)^2}+{1\over
(x_1-x_3)^2(x_2-x_4)^2}+
{1\over(x_1-x_4)^2(x_2-x_3)^2}\right]^{N_{1L}}\cr}}
The factors missing in the expression above are various
contractions
of the polarization tensors. One can implicitly keep track of
these
contractions through the factors of $(x_i-x_j)^{-2}$ (\eg
$(x_1-x_2)^{-2}$ corresponds to a contraction of a pair
of indices between $\xch^{1L}$ and $\xch^{2L}$). The factor
inserted indicates that three of the integration variables are
to be fixed. This removes the $SL(2,R)$ divergence in the full
integration, and provides the Faddeev-Popov determinant.
For the $s$-$t$ channel, we choose: $x_1=0$, $0\le x_2=x\le 1$,
$x_3=1$, $x_4=C\rightarrow\infty$. The above expression then
reduces to
\eqn\factorb{
A_{4,bos}\approx
\int_0^1\!dx\, (1-x)^{p_{2L}\cdot p_{3L}}\, x^{p_{1L}\cdot
p_{2L}}
\left[{1\over x^2}+1+{1\over(1-x)^2}\right]^{N_{1L}}
\left[{1\over(1-x)^2}\right]^{N_{2L}-N_{1L}}
}

Further we have $p_{1L}\cdot p_{2L}=N_{1L}+N_{2L}-2-(s_L/2)$
and $p_{2L}\cdot p_{3L}=2N_{2L}-2-(t_L/2)$. Thus the above
expression \factorb\ becomes a sum of terms of the form
\eqn\integral{
\eqalign{I_{ab}=&\int^1_0 dx (1-x)^{(2N_{1L}-2-2a-t_L/2)}
x^{(N_{1L}+N_{2L}-2-2b-s_L/2)}\cr
=&{\Gamma(2N_{1L}-1-2a-t_L/2)\Gamma(N_{1L}+N_{2L}-1-2b-s_L/2)
\over\Gamma(N_{1L}-N_{2L}+2-2(a+b)+u_L/2)}\cr}}
where $a$ and $b$ are non-negative integers with $a+b\le N_{1L}$.
Considering each of these Gamma function factors in turn:

\smallskip
\noindent{ a) $\Gamma(2N_{1L}-1-2a-t_L/2)$}: From
\mandelstaml, $t_L=O(v^2)$. Hence the argument of this
factor is essentially a positive integer, except for the case
$a=N_{1L}$ (and $b=0$). In the latter case, the argument is
almost $-1$, and this factor in the numerator contributes
an  $O(v^{-2})$ pole. Otherwise this factor is simply a finite
constant.

\smallskip
\noindent{ b) $\Gamma(N_{1L}+N_{2L}-1-2b-s_L/2)$}: From
\mandelstaml,
$s_L/2=(\b+1)^2(N_{1L}-1)+O(v^2)$. Using
$(N_{2L}-1)=\b^2(N_{1L}-1)$, this factor
becomes $-2\beta(N_{1L}-1)+1-2b+O(v^2)$. With $\b\ge1$ and
$N_{1L}\ge2$, this argument is negative for all choices of $a$
and $b$. In fact,
one can show that it is a negative integer up to $O(v^2)$,
and so this factor contributes an $O(v^{-2})$ pole for all values
of $a$ and $b$.

\noindent{ c) $\Gamma(N_{1L}-N_{2L}+2-2(a+b)+u_L/2)$}: From
\mandelstaml, $u_L/2=(\b-1)^2(N_{1L}-1)+O(v^2)$. In this case,
the argument then becomes $-2(\beta-1)(N_{1L}-1)+2-2(a+b)
+O(v^2)$.
Generically, one can show that this argument is a
non-positive integer up to $O(v^2)$, and so this factor
in the denominator contributes a zero of $O(v^{2})$.
\footnote{$^2$}{The only
exception to this behavior is if $\b=1$ (\ie $N_{1L}=N_{2L}$)
and $a=b=0$, in which case, the argument is positive and this
factor no longer contributes a zero. With
further analysis, one finds no $t$-channel pole in this case.
However since both pairs of ingoing and outgoing states
are now identical, one does find an analogous pole corresponding
to massless particle exchange in the $u$-channel where
particles 3 and 4 are interchanged.}

Overall then the generic $I_{ab}$ are finite since the zero in
(c)
cancels the pole in (b). Hence the corresponding terms do not
contribute to the low velocity scattering.
The only relevant term with an $O(v^{-2})$ pole is that with
$(a,b)=(N_{1L},0)$ which yields
\eqn\resulta{\eqalign{
I_{N_{1L},0}=&{\Gamma(-1-t_L/2)\Gamma(N_{1L}+N_{2L}-1-s_L/2)
\over\Gamma(-N_{1L}-N_{2L}+2+u_L/2)}\cr
\approx& {4\beta(N_{1L}-1)\over t_R}{u_R\over s_R}\cr}}
using \mandelstam\ and $t_L=t_R$.
Using $s_R+t_R+u_R=0$, one can rewrite this expression as
\eqn\resultb{
I_{N_{1L},0}\approx-4\beta(N_{1L}-1)\left({1\over t_R}+
{1\over s_R}
\right)}
Note that this contribution included all of the factors of
$(1-x)^2=(x_3-x_2)^2$ in \integral\ so for the polarization
factor in
this term one has only contractions between $\xch^{2L}$ and
$\xch^{3L}$,
and also between $\xch^{1L}$ and $\xch^{4L}$ from the implicit
factors of $(x_4-x_1)^2$.
These are the expected contractions for elastic scattering of
$1\rightarrow4$ and $2\rightarrow3$.

Now gathering up the relevant terms, the total amplitude yields
\eqn\final{\eqalign{
A_{4,het}\approx&2\beta(N_{1L}-1)\,\cc 1 4\,
\cc 2 3\,\left({1\over t_R}+{1\over s_R}\right)\cr
&\ \ \ \left(
{s_R\,t_R\over u_R}\,\zz 1 3\,\zz 2 4
+s_R\,\zz 2 3\,\zz 1 4+t_R\,\zz 1 2\,\zz 3 4\right)\cr
\approx& 2\sqrt{N_{1L}-1}\sqrt{N_{2L}-1}{s_R\over t_R}+
\cdots\cr}}
where we have isolated the only $t$-channel pole.

Note that in this expression, the first factor maybe rewritten
as
$2\sqrt{N_{1L}-1}\sqrt{N_{2L}-1}=m_1m_2$. Now within the low
velocity approximation, one finds $s_R\approx m_1m_2|\vec{v}_1-
\vec{v}_2|^2$. Following \jerome\ in Fourier transforming the
$t$-channel momentum and dividing by $4(m_1m_2)$ to account for
the relativistic normalization of states, one then arrives at a
non-relativistic
potential describing this $t$-channel interaction of the form
\eqn\poten{
U\approx m_1m_2 {|\vec{v}_1-\vec{v}_2|^2\over|\vec{x}_1-
\vec{x}_2|\hphantom{^2}} }
up to numerical prefactors.
This potential then coincides precisely with that describing
the leading order long range interaction of the $a=1$ black
holes.
Thus we have further, dynamical, evidence for the identification
of massive string states with black holes.

\newsec{Discussion}

Following Duff and Rahmfeld's identification of certain
states of the Sen-Schwarz spectrum with extreme black
holes, Sen \senbh\ supported this correspondence by
arguing that string theory could correct the Bekenstein-Hawking
black hole entropy so as to reproduce the logarithm of the
density of states of elementary string states. Peet \peet\
generalized this correspondence as well as the entropy analysis
of Sen to black holes and
massive string states in higher dimensions. In this paper we
presented dynamical evidence for the identification by comparing
two seemingly very different methods of scattering: that
of Manton's metric on moduli space approximation to the
low-velocity dynamics of classical extreme black hole solutions
of the low-energy effective action on the one hand, and the
string four-point amplitude for the scattering of the
corresponding string states on the other.
 The fact that the
results of these two methods are in agreement is rather
remarkable, and points to possible as yet unrealized connections
between classical solutions of string theory and states in
its spectrum.

Our results can be extended to the dynamics of
extremal black holes and corresponding
massive string states in higher dimensions as follows:
Consider the action in $D$ dimensions
\eqn\highdim{
I= {1\over 16\pi G}\int d^D\!x\,\sqrt{-g}\left [R-
{\gamma\over 2}(\partial
\phi)^2-{1\over 4}e^{-a\gamma \phi}F^2 \right]\ ,}
where $\gamma=2/(D-2)$. The same Bogomol'nyi bound \xtremity\
holds in this case. The $\kappa$-symmetric extremal
black holes with $a=\sqrt{D-1}$ scatter trivially \shirscat\
and correspond to string states with $N_R=1/2$ and $N_L=1$
in $D$ dimensions, in analogy with the $a=\sqrt{3}$ case in
$D=4$. The $a=1$ extremal black holes again correspond
to $N_R=1/2$, $N_L>1$ states with $\alpha_L=0$ and the
low-velocity scattering following Manton's method again
matches the corresponding four-point amplitude.

One would also like to extend this analysis to the $N_R=1/2$
string
states with $N_L>1$ and $\a_L\ne0$. These
should correspond to extreme black holes, but a
truncation to an effective action of the form \mbone\ (or
\highdim\ for $D>4$) is not possible. The generalized black hole
solutions are known \refs{\senb,\peet}, but the
Manton metric describing their low-velocity scattering remains
to be calculated. The calculation of the corresponding string
amplitudes
is essentially unchanged, and the final amplitudes still carry
a factor of $2\b(N_{1L}-1)=\b(\a_{1R}^2-\a_{1L}^2)$. Keeping
in mind that the internal momenta are all parallel in these
calculations, one can re-express this factor as $\q_1^TL\q_2$,
where again $L$ is the invariant metric on $O(6,22)$.
Thus the conjectured identification of string states and black
holes
would predict that the low-velocity scattering of these black
holes is governed by an interaction of the form
\eqn\potential{
U\approx \q_1^TL\q_2 {|\vec{v}_1-\vec{v}_2|^{2\hphantom{-3}}
\over|\vec{x}_1-\vec{x}_2|^{D-3}} }
in $D$ dimensions.

One possible gap in the correspondence tested here
is that there seem to be many
string states corresponding to a single extreme
black hole. While the black hole solution in the low energy
theory are completely fixed once the charge vector is specified,
one still has the freedom to specify the polarization tensor for
the string states.
One possible way around this problem is, following Sen
\senbh, to claim that quantum corrections to the entropy of
the black hole would correspond to the entropy of the string
states. An alternate suggestion
is that when one goes beyond the massless fields in
effective action to include
the massive string states and Kaluza-Klein modes of the full
theory, one will find new black hole solutions such that there
will be a one-to-one correspondence between the black holes
and string states. At present,
this suggestion has yet to be investigated.
A related problem worth considering is the
conversion between different polarizations of elementary string
states by interactions with the moduli scalars. This process
should
correspond to the conversion between different extremal black
holes,
in analogy with the conversion of monopoles into dyons found in
the study of BPS monopoles \atiyah.

In comparing the dynamics of black holes and massive string
states, we must be careful to remain within the validity
of the low-velocity approximation. For example, the metric
on moduli space for two $a=1$ black holes seems to indicate
that such black holes will never coalesce \shirscat.
Examining the string amplitude \final\ though, one does
find that there are $s$-channel poles. These
arise from the two incoming particles merging to form a
single string state with mass $m=m_1+m_2=(1+\b)m_1$ and internal
momenta $\a_{R}=\a_{1R}+\a_{2R}=(1+\b)\a_{1R}$ and $\a_L=0$.
Therefore the intermediate state still satisfies $m^2=\a_{R}^2$
and hence is still a state with $N_R=1/2$, $N_L>1$ and
$\a_L=0$. Thus the intermediate state should also correspond to
another $a=1$ black hole. The string theory calculation then
seems to indicate that $a=1$ black holes can in fact merge,
in apparent contradiction to the Manton scattering results.
However, following the discussion of \fere, when the relative
separation
$r\to 0$, the validity of Manton's method breaks down and
so in fact there is no contradiction.

Finally, we note that the agreement between a flat metric for
the scattering of elementary string solutions and a vanishing
four-point amplitude for string winding states in the infinite
winding radius limit was found previously in \macstr\ in order
to provide dynamical evidence for the identification of
elementary string solutions of the equations
of motion with macroscopic fundamental winding states.
The present comparison for $a=\st$ extremal black holes
involves point-like string states
and black holes and so requires a different limit. It is
interesting that the same phenomenon of trivial scattering,
from both viewpoints, appears in the two different scenarios.

\vskip1truecm

\noindent
{\bf Acknowledgements:}

\noindent
We would like to thank Mike Duff for suggesting this
problem and for helpful discussions This research was supported
by NSERC of Canada and Fonds FCAR du Qu\'ebec.

\vfil\eject
\listrefs
\bye